\documentclass[11pt,a4paper]{article}

\usepackage{amsmath,amssymb,graphicx,cite}

\begin{document}

\title{Shadows of rotating black holes in alternative theories}

\author{Leonardo Amarilla and Ernesto F. Eiroa$^{1,2,}$\thanks{e-mail: eiroa@iafe.uba.ar}\\
{\small $^1$ Instituto de Astronom\'{\i}a y F\'{\i}sica del Espacio (IAFE, CONICET-UBA),}\\
{\small Casilla de Correo 67, Sucursal 28, 1428, Buenos Aires, Argentina}\\
{\small $^2$ Departamento de F\'{\i}sica, Facultad de Ciencias Exactas y 
Naturales,} \\ 
{\small Universidad de Buenos Aires, Ciudad Universitaria, 1428, 
Buenos Aires, Argentina}} 

\maketitle

\begin{abstract}
We briefly review some recent advances in the study of the shadows of rotating black holes in alternative theories. The size and the shape of the shadow depend on the mass and the angular momentum, and they can also depend on other parameters specific of the particular model adopted. As an example, we show the results corresponding to a rotating braneworld black hole. \\


\end{abstract}

\section{Basic aspects}

Part of the photons emitted from a luminous background behind a black hole, i.e. those with small impact parameters, end up falling into the black hole, not reaching the observer, and producing a  completely dark zone denominated the shadow. The apparent shape of a black hole is thus defined by the boundary of the shadow. The shadow of a non-rotating black hole is circular, but the presence of spin produces a deformation\cite{chandra92,ve04,hm09}, because co-rotating photons interact with a more intense potential than counter-rotating ones, resulting in a closer approach to the black hole. This topic has been recently investigated by several researchers, both in Einstein gravity and in modified theories\cite{ss09,aeg10,ae12,ae13,gpl14}, with the confidence that direct observation of supermassive black holes in the center of our galaxy and also in nearby galaxies will be feasible in the near future. 

We begin by resuming the procedure for obtaining the contour of the shadow for the Kerr black hole\cite{chandra92}. In Boyer-Lindquist coordinates, the Kerr metric can be written in the form (G=c=1)
\begin{equation}
ds^2  =  -\frac{\Delta}{\Sigma} \left(dt - a \sin^2\theta d\varphi \right)^2 + \Sigma \left(\frac{dr^2}{\Delta} + d\theta^{2}\right) + \frac{\sin^2\theta}{\Sigma}\left[ a dt - (r^2+a^2) d\varphi \right]^2 ,
\label{metric}
\end{equation}
where 
\begin{equation}
\Sigma=r^{2}+a^{2}\cos^{2}\theta,  \;\;\;\;\; \Delta=r^{2}-2Mr+a^{2},
\label{sdkerr}
\end{equation}
with $M$ the mass and $a=J/M$ the rotation parameter. The horizons, with radii $r_{\pm}=M^{2} \pm \sqrt{M^{2}-a^{2}}$, are obtained from the equation $\Delta =0$; $r_{+}$ is the event (outer) horizon and $r_{-}$ is a Cauchy (inner) horizon. Kerr spacetime is stationary and axisymmetric. These symmetries have associated Killing vectors, so $p_{t}=-E$ y $p_{\varphi}=L_{z}$ are conserved along the geodesic movement of a particle ($E$ is the energy and $L_z$ is the axial component of the angular momentum). There is an additional hidden symmetry corresponding to another conserved quantity, the Carter constant\cite{chandra92} $\mathcal{K}$. The geodesics are determined from the Hamilton-Jacobi equation
\begin{equation}  \label{SHJ1}
\frac{\partial S}{\partial \lambda}=-\frac{1}{2}g^{\sigma\nu}\frac{\partial S}{\partial x^{\sigma}}\frac{\partial S}{\partial x^{\nu}},
\end{equation}
where $\lambda$ is an affine parameter and $S$ is the Jacobi action. This action can be separated for the Kerr geometry in the form\cite{chandra92}
\begin{equation}  \label{SHJ2}
S=\frac{1}{2}m ^2 \lambda - E t + L_z \varphi + S_{r}(r)+S_{\theta}(\theta),
\end{equation}
where $m$ is the particle mass. In the case of photons ($m =0$), the equations of motion resulting from $p_{\nu}=\partial S / \partial x^{\nu}$  have the form\cite{chandra92}
\begin{equation}
\Sigma\frac{dt}{d\lambda }=\frac{1}{\Delta}\left[E(r^{2}+a^{2})^{2}-\Delta a^{2} E\sin^{2}\theta-2MarL_z\right], 
\end{equation}
\begin{equation}
\Sigma\frac{dr}{d\lambda }=\pm \sqrt{\mathcal{R}}, 
\end{equation}
\begin{equation}
\Sigma\frac{d\theta}{d\lambda }=\pm \sqrt{\Theta}, 
\end{equation}
\begin{equation}
\Sigma\frac{d\varphi}{d\lambda }=\frac{1}{\Delta}\left[2MEar+(\Sigma-2Mr)L_z \csc^{2}\theta  \right],
\end{equation}
where 
\begin{equation}
\mathcal{R}(r)=\left[(r^2+a^2)E -a L_z \right]^2-\Delta\left[\mathcal{K}+(L_z -a E)^2\right], \label{R}
\end{equation}
\begin{equation}
\Theta(\theta)=\mathcal{K}+\cos^2\theta\left[a^2E^2-L_{z}^{2}\csc^2\theta \right].
\label{Theta}
\end{equation}
The unstable orbits of photons with constant $r$ should fulfill the conditions $\mathcal{R}=0$ and $d\mathcal{R}/dr=0$. One can establish a relation between this radius and the conserved quantities, so the system of equations can be solved for the impact parameters $\xi=L_z/E$ and $\eta=\mathcal{K}/E^2$ corresponding  to the unstable photon orbits to give
\begin{equation}\label{xietak1}
\xi (r)=\frac{(r^{2}-a^{2})M - \Delta r}{a(r-M)},
\end{equation}
\begin{equation}\label{xietak2}
\eta (r) =\frac{r^{3}\left[4 M \Delta - r (r-M)^{2}\right]}{a^{2}(r-M)^{2}}.
\end{equation}
The apparent position of the photon sphere in the sky of  a distant observer, produces the contour of the shadow. To describe it, we adopt the celestial coordinates\cite{ve04}:
\begin{equation}  \label{alphabeta1}
\alpha=\lim_{r_{0}\rightarrow \infty}\left( -r_{0}^{2}\sin\theta_{0}\frac{d\varphi}{dr}\right)
\;\;\;\;\;
\mathrm{and}
\;\;\;\;\;
\beta=\lim_{r_{0}\rightarrow \infty}r_{0}^{2}\frac{d\theta}{dr},
\end{equation}
where $r_{0}$ goes to infinity because we consider an observer far away from the black hole, and $\theta_{0}$ is the angular coordinate of the observer, i.e. the inclination angle between the rotation axis of the black hole and the line of sight of the observer. The coordinates $\alpha $ and $\beta$ are the apparent perpendicular distances of the image as seen from the axis of symmetry and from its projection on the equatorial plane, respectively. They take the form \cite{ve04}
\begin{equation}  \label{alphabeta2}
\alpha=-\xi\csc\theta_{0}
\;\;\;\;\;
\mathrm{and}
\;\;\;\;\;
\beta=\pm \sqrt{\eta + a^{2}\cos ^{2}\theta_{0}-\xi^{2}\cot ^{2}
\theta_{0}}.
\end{equation}
For an observer located in the equatorial plane of the object ($\theta_{0}=\pi/2$), natural in the case of the Galactic supermassive black hole, Eqs. (\ref{alphabeta2}) simplify to 
\begin{equation}  \label{alphabeta3}
\alpha=-\xi
\;\;\;\;\;
\mathrm{and}
\;\;\;\;\;
\beta=\pm \sqrt{\eta}. 
\end{equation}
The effects of rotation  are stronger when seen from this plane. 

\begin{figure}[t!]
\begin{center}
\includegraphics[width=0.25\textwidth]{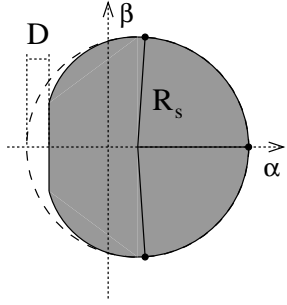}
\end{center}
\caption{Schematic plot of the shadow of a rotating black hole.}
\label{fig1}
\end{figure}

The shadow can be characterized by defining observables. Here, we adopt\cite{hm09} two parameters: the radius of a reference circle $R_s$ passing by three points of the shadow, the top position $(\alpha_t, \beta_t)$, the bottom position $(\alpha _b,\beta _b)$, and the point corresponding to the unstable retrograde circular orbit seen by an observer on the equatorial plane $(\alpha _r,0)$; and the distortion parameter $\delta _{s}=D/R_s$, with $D$ the difference between the endpoints of the circle and of the shadow, both of them at the opposite side of the point $(\alpha _r,0)$. A schematic diagram is shown in Fig. \ref{fig1}. The radius $R_s$ measures the overall size of the shadow and $\delta_s$ gives its deformation with respect to the reference circle. These observables are given by $R_{s}=[(\alpha _t -\alpha_r)^2 + \beta_t ^2][2|\alpha _t -\alpha_r|]^{-1}$ and $\delta _s=(\tilde{\alpha}_p - \alpha_p)R_{s}^{-1}$, where $(\tilde{\alpha}_p, 0)$ and $(\alpha_p, 0)$ are the points where the reference circle and the contour of the shadow cut the horizontal axis at the opposite side of $(\alpha_r, 0)$, respectively.

The deformation of the shadow grows with the inclination angle $\theta _0 $ of the observer with respect to the rotation axis of the black hole. For a polar observer ($\theta_{0}=0$ or $\theta_{0}=\pi$) there is no deformation, while for one in the equatorial plane ($\theta_{0}=\pi/2$) the gravitational effects on the shadow are larger. If there are other parameters associated to the black hole, e.g. the charge, it will depend on them too, in a complex way. In the case of black holes for which the corresponding Hamilton-Jacobi equation can be separated in the form (\ref{SHJ2}), a similar method can be applied to find the contour of the shadow.  The first step is to separate the Hamilton-Jacobi action for the null geodesics into a radial and an angular part. The null geodesics are then parametrized in terms of the conserved quantities $\xi =L_z/E$ and $\eta =\mathcal{K}/E^2$. The boundary of the shadow is determined by the geodesics of photons with parameters $\xi $ and $\eta $ corresponding to the unstable spherical orbits around the deflector. Then, the quantities $\xi $ and $\eta $ can be related with the celestial coordinates $\alpha $ and $\beta$. Finally, the observables can be found in terms of the relevant parameters of the black hole geometry under study.

\section{Example: braneworld rotating black hole}

As an example of application of the method to alternative theories, we analyze the shadow of a braneworld black hole in the Randall-Sundrum scenario\cite{ss09,ae12}. We consider the geometry\cite{ag05} given by Eq. (\ref{metric}), with the same $\Sigma$ but now $\Delta$ replaced by
\begin{equation}
\Delta _{b}= r^2 + a^2 -2 M r+ Q,
\label{dtidal}
\end{equation}
where $Q$ is the tidal charge. This spacetime was obtained under the assumption that the induced geometry on the 3D brane has the Kerr-Schild form \cite{ag05}. It is not known if this choice of the metric on the brane is indeed fulfilled by an exact bulk metric  \cite{ag05}. The tidal charge is interpreted as an imprint of the gravitational effects from the bulk space; it can take any sign, but some researches think that $Q<0$ is more natural \cite{ag05}. When $Q=0$ one recovers the Kerr geometry.  The event horizon radius is $r_{h}= M + \sqrt{M^2 - a^2- Q}$ if $Q  \leq Q_c =M^2 -a^2$ (in other case there is a naked singularity). When $Q>0$, this condition leads to $|a|\le M$; but for $Q<0$ the maximally rotating black hole has $|a|>M$. 

The Hamilton-Jacobi equation (\ref{SHJ1}) determines the geodesics. As in Kerr case, the problem is separable, so the Jacobi action $S$ can be written in the form (\ref{SHJ2}). For null geodesics ($m =0$), from Eq. (\ref{SHJ1}) one can obtain the equations of motion:
\begin{equation}
\Sigma\frac{dt}{d\lambda}=a(L_z -aE\sin^2\theta)+
\frac{r^2+a^2}{\Delta_{b}}\left[(r^2+a^2)E -a L_z \right],
\end{equation}
\begin{equation}
\Sigma\frac{dr}{d\lambda}=\pm \sqrt{\mathcal{R}},
\end{equation}
\begin{equation}
\Sigma\frac{d\theta}{d\lambda}=\pm \sqrt{\Theta},
\end{equation}
\begin{equation}
\Sigma\frac{d\varphi}{d\lambda}= L_z \csc^2\theta -aE + 
\frac{a}{\Delta_{b}}\left[(r^2+a^2)E -a L_z \right],
\end{equation}
where 
\begin{equation}
\mathcal{R}(r)=\left[(r^2+a^2)E -a L_z \right]^2-\Delta_b \left[\mathcal{K}+(L_z -a E)^2\right], \label{Rbw}
\end{equation}
\begin{equation}
\Theta(\theta)=\mathcal{K}+\cos^2\theta\left[a^2E^2-L_{z}^{2}\csc^2\theta \right],
\label{Thetabw}
\end{equation}
with $\mathcal{K}$ the Carter constant. The expressions for $\mathcal{R}(r)$ and $\Theta(\theta)$ have the same form as in Kerr geometry, with $\Delta$ replaced by $\Delta _{b}$. The spacetime is asymptotically flat, so the trajectory of photons are straight lines at infinity. The orbits with constant $r$  satisfying $\mathcal{R}(r)=0=d\mathcal{R}(r)/dr$, determine the values of the impact parameters $\xi=L_z/E$ and $\eta=\mathcal{K}/E^{2}$ associated to the contour of the shadow:
\begin{equation}
\xi(r)=\frac{a^{2}(M+r)+r\left[ (r-3M)r+2 Q \right] }{a(M-r)},
\label{eqxieta1}
\end{equation}
\begin{equation}
\eta(r)=\frac{r^{2}}{a^{2}(r-M)^{2}}\left\{ 4a^{2}(Mr-Q)-\left[ r(r-3M)+2 Q \right] ^{2}\right\} .
\label{eqxieta2}
\end{equation}
By taking $Q=0$ the corresponding values for Kerr geometry are recovered, given by Eqs. (\ref{xietak1}) and  (\ref{xietak2}). If we calculate $d\varphi/dr$ and $d\theta/dr$ and take the limit of a far away observer, we obtain that the celestial coordinates have the same form as for the Kerr metric, i.e. Eqs. (\ref{alphabeta2}),  with the new $\xi $ and $\eta$ given by Eqs. (\ref{eqxieta1}) and (\ref{eqxieta2}). 

\begin{figure}[t!]
\begin{center}
\includegraphics[width=0.31\linewidth]{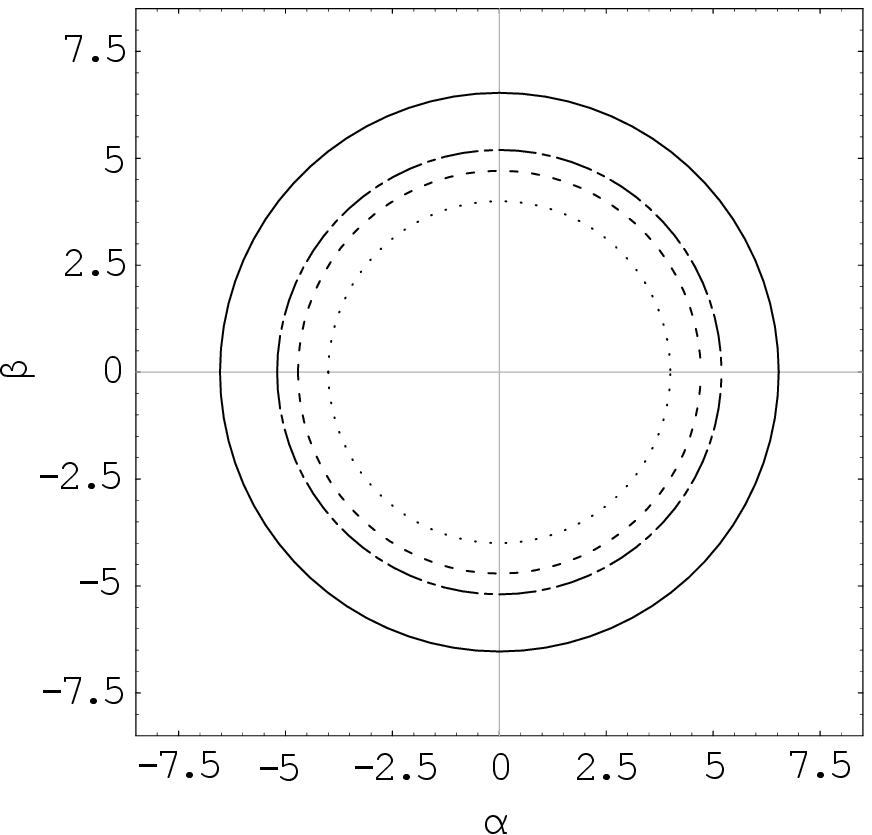}
\hspace{0.01\linewidth}
\includegraphics[width=0.31\linewidth]{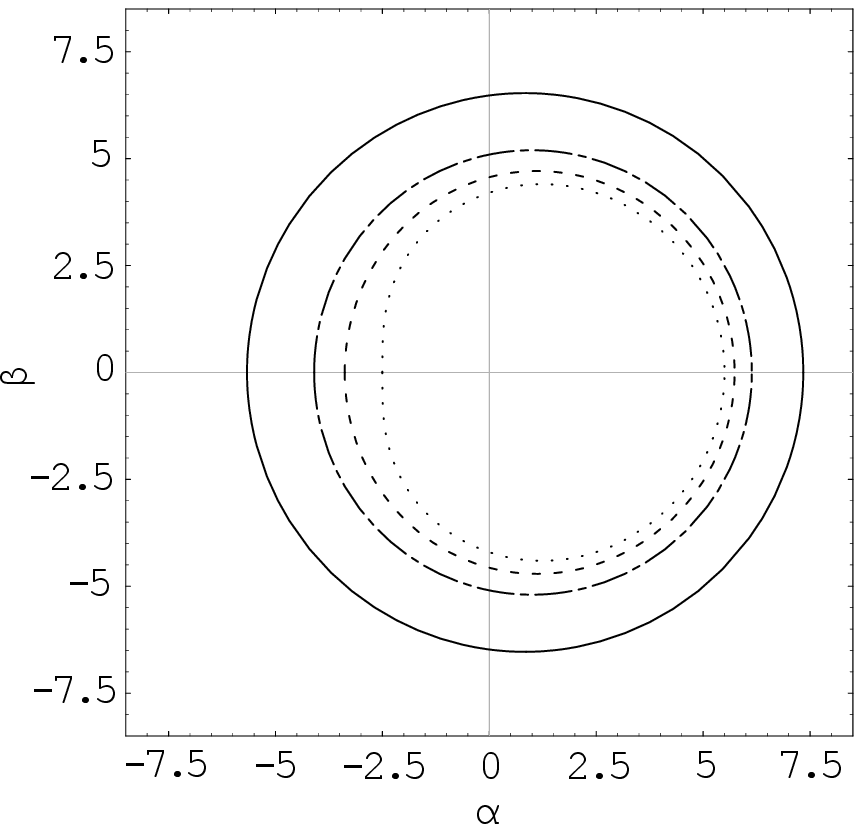}
\hspace{0.01\linewidth}
\includegraphics[width=0.31\linewidth]{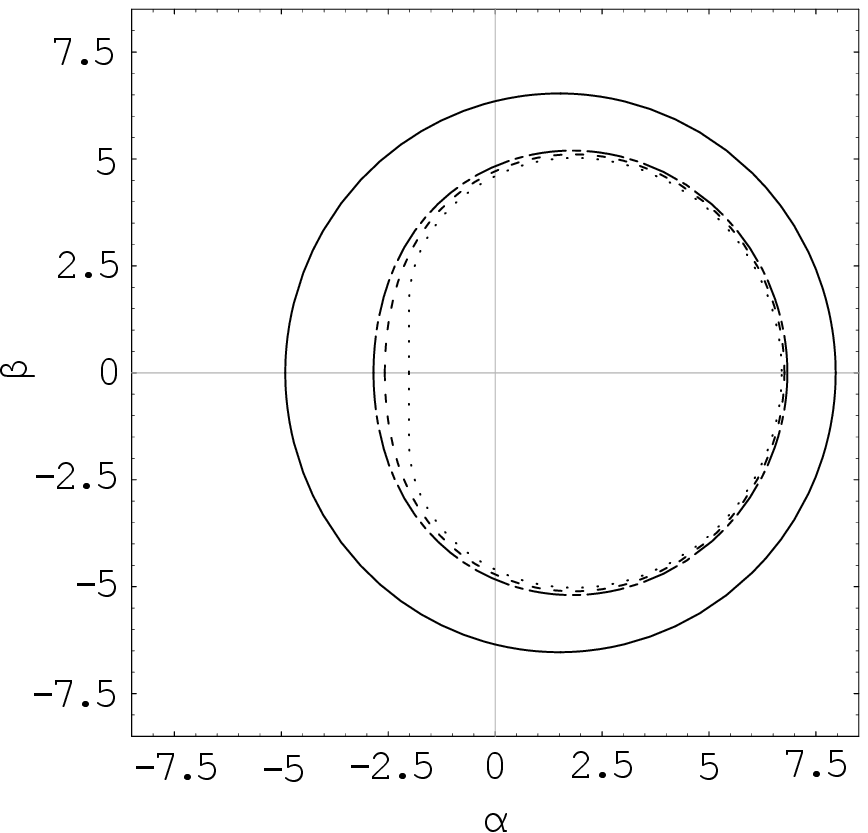}
\end{center}
\vspace{-0.4cm}
\caption{Contour of the shadow cast by a braneworld black hole, for $\theta _0=\pi /2$. Left: $a/M=0$, $Q/M= -2$ (full line), $Q/M=0$ (dashed-dotted line), $Q/M=0.5$ (dashed line), and $Q_{c}/M=1$ (dotted line). Center:  $a/M=0.5$,   $Q/M= -2$ (full line), $Q/M=0$ (dashed-dotted line), $Q/M=0.5$ (dashed line), and $Q_{c}/M=0.75$ (dotted line).  Right:  $a/M=0.9$,   $Q/M= -2$ (full line), $Q/M=0$ (dashed-dotted line), $Q/M=0.1$ (dashed line), and $Q_{c}/M=0.19$ (dotted line).}
\label{fig2}
\end{figure}

\begin{figure}[t!]
\begin{center}
\vspace{0.4cm}
\includegraphics[width=0.3\linewidth]{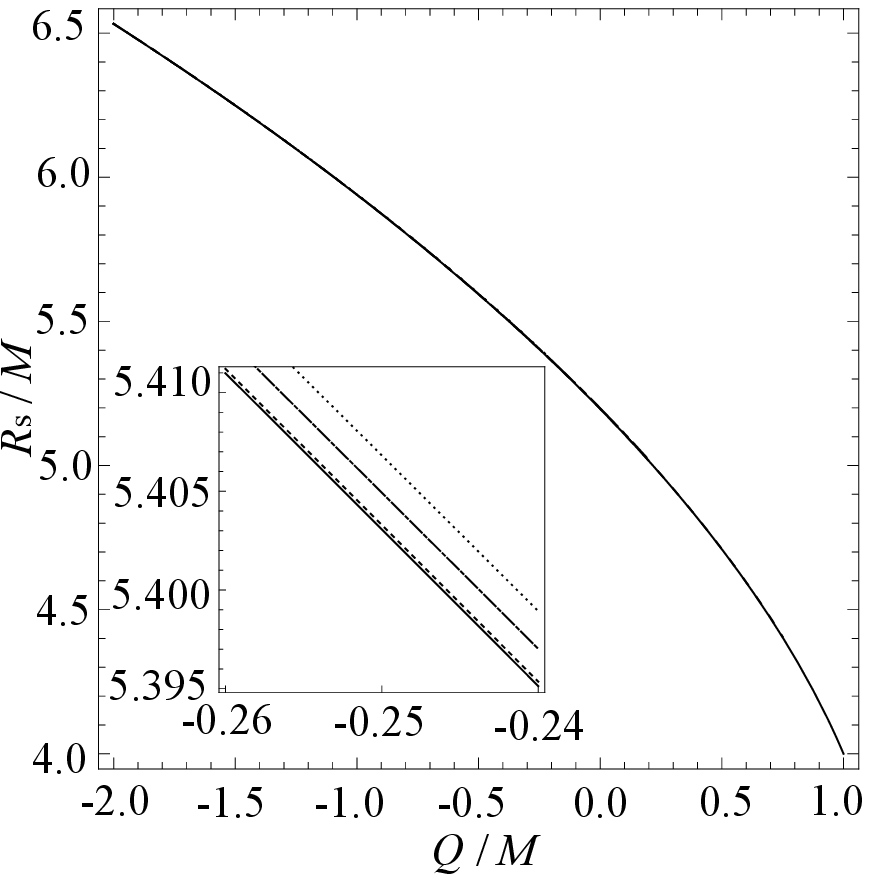}
\hspace{0.01\linewidth}
\includegraphics[width=0.31\linewidth]{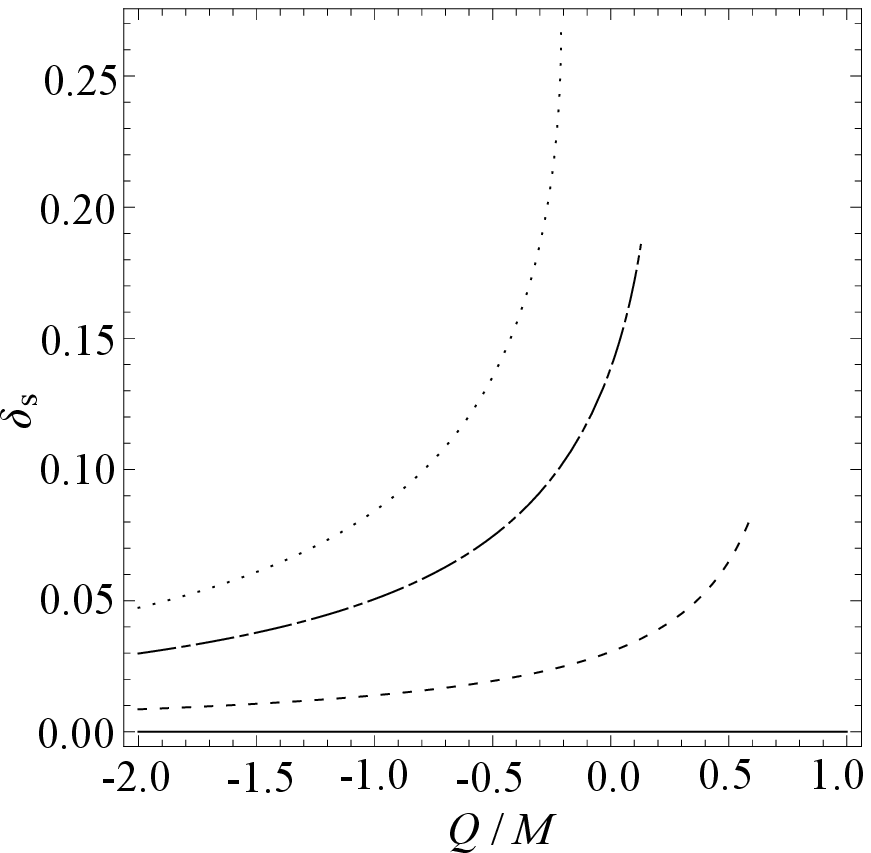}
\hspace{0.01\linewidth}
\includegraphics[width=0.31\linewidth]{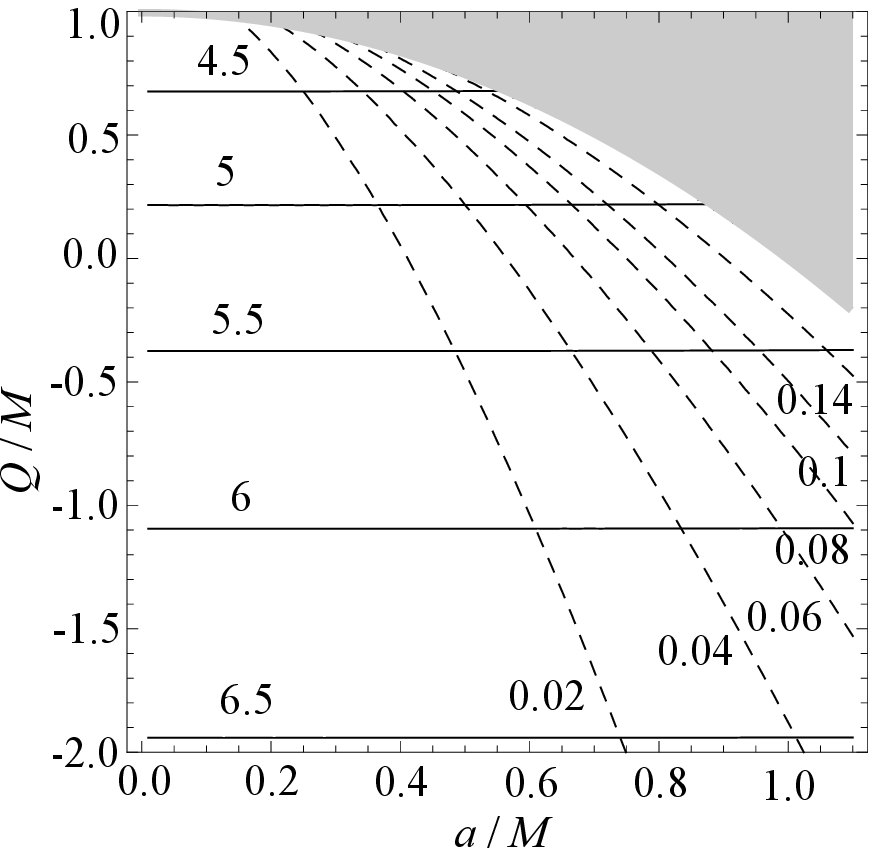}
\end{center}
\vspace{-0.4cm}
\caption{Observables $R_{s}/M$ (left) and $\delta _{s}$ (center) as functions of $Q/M$, for $\theta _0=\pi /2$, and $a/M=0$ (full line), $a/M=0.5$ (dashed line), $a/M=0.9$ (dashed-dotted line), and $a/M=1.1$ (dotted line). Contour curves (right) of constant $R_s/M$ (full line) and $\delta _{s}$ (dashed line) in the plane $(a/M,Q/M)$, for $\theta _0=\pi /2$ (the gray zone corresponds to naked singularities). }
\label{fig3}
\end{figure}

In Fig. \ref{fig2} the contour of the shadow is plotted for some representative values of the parameters. In Fig. \ref{fig3}, the observables $R_{s}/M$ and $\delta_{s}$ are shown as functions of $Q/M$. From the plot corresponding to $R_{s}/M$, we see that the size of the shadow for a fixed value of $a/M$ decreases with $Q/M$, resulting in a larger shadow than in Kerr geometry for $Q<0$ and a smaller shadow for $Q>0$. For fixed $Q/M$, the difference in the values of $R_{s}/M$ for $a/M$ in the range between $0$ and $1.1$ is of order $10^{-3}$, leading to a small variation in the size of the shadow as a function of $a/M$. The deformation of the shadow, characterized by $\delta_{s}$, increases with $Q/M$, so a less distorted silhouette is obtained for $Q<0$ and a more deformed one for $Q>0$. For fixed $Q/M$, the deformation of the shadow increases with $a/M$. If $M$ and $\theta _{0}$ are independently known, measurements of $R_{s}$ and $\delta _{s}$ could serve to find $a$ and $Q$. 

The angular size of the shadow can be estimated by $\theta _s = R_{s}/D_{o}$, with $D_{o}$ the distance between the black hole and the observer. In the case of the supermassive black hole at the Galactic center\cite{guil09}, with mass $M=4.3 \times 10^{6}M_{\odot}$ and located at a distance of $D_{o}=8.3$ kpc, for $\theta_0=\pi /2$, we can see that a resolution much better than $1 \, \mu \mathrm{as}$ is needed in order to find differences with the Kerr shadow\cite{ae12}. Similar remarks can be done for other rotating black holes in alternative theories\cite{aeg10,ae13}.

\section{Discussion}

When the Hamilton-Jacobi equation can be separated, the contour of the shadow can be found by following a standard procedure. The analysis of the shadow will be an useful tool for obtaining properties of astrophysical black holes and comparing gravitational theories. Direct imaging of black holes will be possible in the next years\cite{mmg12,joh12,lu14}. The Event Horizon Telescope, consisting of a very long baseline interferometry network of radio-telescopes scattered over the Earth operating at millimeter and sub-millimeter wavelengths ($230-450$ GHz), will reach an angular resolution of $15$ $\mu$as or better (depending on the number of stations), enough to observe the shadow of the supermassive Galactic black hole and those corresponding to nearby galaxies. The projected space-based Millimetron mission is expected to provide the angular resolution of $0.3$ $\mu$as or less at $0.4$ mm. There are also proposals of space facilities with a high resolution ($0.1$ $\mu$as) in X-rays  like MAXIM. However, more advanced instruments with a higher resolution are necessary for detecting deviations from General Relativity like those considered here.

\section*{Acknowledgments}

This work was supported by CONICET and UBA.

\end{document}